\documentclass[conference,10pt]{IEEEtran}
\IEEEoverridecommandlockouts
\usepackage{amsmath,amssymb,amsfonts}
\usepackage{algorithmic}
\usepackage{graphicx}
\usepackage{textcomp}
\usepackage{xcolor}
\usepackage{caption}
\usepackage{subcaption}
\usepackage{tikz}
\usepackage{lipsum}
\usepackage{float}

\def\BibTeX{{\rm B\kern-.05em{\sc i\kern-.025em b}\kern-.08em
    T\kern-.1667em\lower.7ex\hbox{E}\kern-.125emX}}
    
\usepackage[
backend=biber,
style=ieee,
sorting=none
]{biblatex}
\newcommand\copyrighttext{%
  \footnotesize  This paper has been accepted for publication in IEEE Conference on Dependable and Secure Computing, Edinburgh, UK, June 2022. This is an author’s copy, The respective copyrights are with IEEE.
}
\newcommand\copyrightnotice{%
\begin{tikzpicture}[remember picture,overlay]
\node[anchor=south,yshift=10pt] at (current page.south) {\fbox{\parbox{\dimexpr\textwidth-\fboxsep-\fboxrule\relax}{\copyrighttext}}};
\end{tikzpicture}%
}

\begin{document}

\title{Graph Neural Network-based Android Malware Classification with Jumping Knowledge\\
}

\author{\IEEEauthorblockN{Wai Weng Lo\IEEEauthorrefmark{1},
Siamak Layeghy\IEEEauthorrefmark{2},
Mohanad Sarhan\IEEEauthorrefmark{3}, 
Marcus Gallagher\IEEEauthorrefmark{4}, and \\
Marius Portmann\IEEEauthorrefmark{5}}

\IEEEauthorblockA{School of Information Technology and Electrical Engineering\\
The University of Queensland, Brisbane, Australia}

\IEEEauthorblockA{Email: \IEEEauthorrefmark{1}w.w.lo@uq.net.au,
\IEEEauthorrefmark{2}siamak.layeghy@uq.net.au,
\IEEEauthorrefmark{3}m.sarhan@uq.net.au,
\IEEEauthorrefmark{4}marcusg@itee.uq.edu.au,
\IEEEauthorrefmark{5}marius@itee.uq.edu.au}
}
\maketitle

\begin{abstract}
This paper presents a new Android malware detection method based on Graph Neural Networks (GNNs) with Jumping-Knowledge (JK). Android function call graphs (FCGs) consist of a set of program functions and their inter-procedural calls. Thus, this paper proposes a GNN-based method for Android malware detection by capturing meaningful intra-procedural call path patterns. In addition, a Jumping-Knowledge technique is applied to minimize the effect of the over-smoothing problem, which is common in GNNs. The proposed method has been extensively evaluated using two benchmark datasets. The results demonstrate the superiority of our approach compared to state-of-the-art approaches in terms of key classification metrics, which demonstrates the potential of GNNs in Android malware detection and classification.
\end{abstract}


\section{Introduction}
Android is the most used widely mobile operating system with 73.0\% \cite{noauthor_mobile_nodate} of the smartphone market share in November 2021. Due to its open-source nature, Android provides flexibility for mobile software developers to create custom Android Application Packages (APKs). However, this has been an effective method for cybercriminals to create malicious applications to access user sensitive information such as credit cards and contact information. In contrast to other closed source platforms such as Apple iOS, hackers can inspect the application's source code to develop exploits \cite{faruki2014android}. Moreover, users can install untrusted third-party applications on Android devices, allowing hackers to distribute the malware efficiently. Therefore, mobile anti-malware solutions are critical for the detection and prevention of mobile malware. 

\copyrightnotice

Traditional anti-malware solutions are mainly based on signature-based detection techniques \cite{gandotra2014malware}, which rely on the analysis and comparison of malware attack signatures to a list of pre-identified signatures. However, this method of traditional detection methods cannot effectively detect unknown malware variants, such as zero-day malware. In contrast, Machine Learning (ML)-based \cite{gandotra2014malware} anti-malware approaches can detect unknown malware variants, using both static and dynamic features such as API calls, permissions \cite{huang2019deep}\cite{arp2014drebin}. In recent years, anti-malware vendors have tried to apply new ML models and techniques, particularly deep learning-based approaches, to develop new anti-malware solutions.

\begin{figure}[t]
    \centering
        \includegraphics[width=0.75\columnwidth]{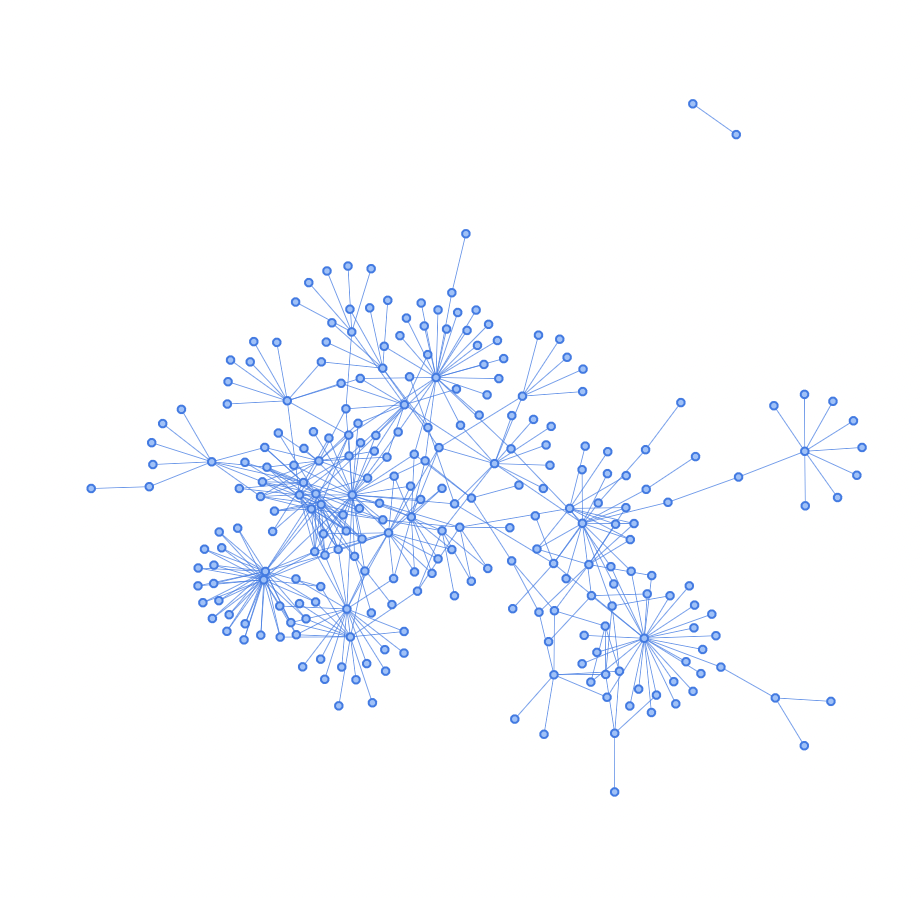}
        %\label{rfidtest_yaxis}
    \caption{Example of SMS malware's FCG}
    \label{fig:API}
\end{figure}

Graph Neural Networks (GNNs) \cite{wu2020comprehensive} represent one of the most recent and fastest growing areas in Machine Learning. Their power to capture topological patterns of graph-based data can be applied in many real-world applications, such as social media networks, biology, telecommunications, etc. Android Function Call Graphs (FCGs) consist of a set of program functions and their inter-procedural calls, which can be extracted from Android Application Package (APK) files and represented as a graph. The corresponding graph structures can be utilised for Android malware detection based on graph representation learning. As an illustration, Figure \ref{fig:API} shows the example of the FCG of SMS malware.

The main problem of ML algorithms \cite{arp2014drebin}\cite{li2018significant}\cite{wang2018detecting}\cite{bai2020famd} is that they require domain knowledge from experts to extract different types of features which is very complex and time-consuming. Moreover, those ML methods have not considered the chain reaction between different function calls, which should be considered. Therefore, the main motivation of this paper is to perform automatic Android malware detection and classification by using GNNs based on FCGs without the use of any handcrafted features. 

GNNs are susceptible to the oversmoothing problem\cite{li2019deepgcns} and their performance can degrade significantly with an increase in the number of neural network layers. To address this problem, we applied the Jumping Knowledge (JK) technique, which combines intermediate representations and jumping to the last layers. Our results indicate that the proposed method can outperform the state-of-the-art approaches in terms of key classification metrics, which demonstrates the potential of GNNs in Android malware classification, and provides motivation for further research.

In summary, the key contributions of this paper are:

\begin{itemize}
\setlength\itemsep{0.8em}
\item The design of an Android malware detection and classification system using a GNN model with JK technique to capture the topological information embedded in Function Call Graphs. A key benefit of the approach is that it does not require handcrafted feature extractions from domain experts.

\item The comprehensive evaluation of the proposed framework using two benchmark datasets demonstrates the superiority via comparison state-of-the-art approaches.
\end{itemize}

\section{\uppercase{Related Works}}
% \subsection{Android Malware Classification based on Machine Learning}
\subsection{Shallow Learning-based Android Malware Classification}
Daniel et al. \cite{arp2014drebin} proposed an explainable ML approach to classify Android malware based on linear Support Vector Machines (SVMs). This approach requires several features from the APKs, including permissions and sensitive APIs to identify malware. The extracted feature sets are projected into a joint vector space for the linear SVM to detect the malware and explain the results.
In \cite{yerima2013new}, a Bayesian classifier was applied to detect Android malware by using 58 defined code-based attributes. The set of 1,000 non-malicious apps and 1,000 malicious apps from 49 different families were used to evaluate the classifier. In \cite{jerome2014using},  opcode-sequence n-gram features were used for Android malware classification. The n-gram features were used to train a classifier to distinguish between benign and malicious APKs. 
However, this approach is not robust to bytecode-level obfuscation as the opcode sequences can be directly altered.  

In \cite{wang2018detecting}, the authors extracted 11 types of static features from each app and employed an ensemble classifier which consists of Support Vector Machine (SVM), K-Nearest Neighbor (KNN), Naive Bayes (NB), Classification and Regression Tree (CART) and Random Forest (RF) classifiers to distinguish malicious from benign APKs. 

Li et al. \cite{li2018significant} proposed SIGPID, which is based on permission usage to detect Android malware. They mined the permission data using 3-levels of pruning techniques to identify the most 22 significant permissions as features. A Support Vector Machine (SVM) approach was used to detect malicious apps. Xu et al. \cite{xu2016iccdetector} extract Inter Component Communication (ICC) patterns from the source code to recognize malware that utilizes inter-component communication to launch stealthy attacks.

\subsection{Deep Learning-based Android Malware Classification}
McLaughlin et al. \cite{mclaughlin2017deep} applied convolution neural networks (CNNs) on the raw opcode sequences for detecting Android malware. The opcode sequences were extracted from disassembled apps. Then, the extracted opcode sequences were encoded as one-hot vectors and fed to an opcode embedding layer for training the CNNs. \cite{khan2019analysis}\cite{wai}\cite{he} convert malware samples as grayscale images then fed them to the CNNs for malware detection and classification. 

In \cite{vinayakumar2017deep}, Android permission sequences were extracted into features through word embedding using an LSTM model. The word embedding was fed into a fully connected layer with a Sigmoid activation function for malware detection. In \cite{huang2019deep}, the API features were mapped to the hand-refined graph matrix as input for the CNN classifier. The API feature selection was based on API occurrence frequency between benign and malware. The results demonstrated that the top 20 API calls can achieve a high F1 score of 94.3\% in Android malware detection. Bai et al. \cite{bai2020famd} applied a Fast Correlation-Based Filter (FCBF) on the n-grams of opcodes in order to reduce feature dimensionality and perform malware detection.

\subsection{Android Malware Detection based on Graph Representation Learning}
In \cite{hashemi2017graph}, the authors generated OpCode graphs from the execution files and used the Power Iteration method to embed the graph into a low dimensional feature space, to serve as input for ML algorithms such as k-nearest neighbour (KNN) and support vector machine (SVM) to perform malware classification. A similar approach was used by Hashem et al. \cite{pektacs2020deep} who applied graph embedding techniques for Android malware classification. The API call graphs were transformed into a lower dimensional space using graph embedding techniques such as HOPE \cite{ou2016asymmetric} and SDNE \cite{wang2016structural}.  The extracted graph embedding was then fed to an 1D CNN \cite{zhang2015character} for Android malware detection. However, these proposed methods are shallow encoders that cannot incorporate and leverage node features of the graphs.  In \cite{zhang2014semantics}, DroidSIFT was proposed for Android malware classification based on the weighted contextual API dependency graph and the classifier achieved  93\% accuracy on the Genome dataset \cite{zhou2012dissecting}.

%--->>

In \cite{gao2021gdroid}, GDroid was proposed for Android malware detection by utilizing word embedding and GNN techniques. The skip-gram model extracted the features for graph nodes based on API sequences. The extracted node features were mapped to the heterogeneous graph to form APP to API relationships to train the GNNs to detect malicious app nodes. However, the proposed method is transductive, which cannot generalize zero-day malware and unknown applications, as we cannot expect every API call to exist in the skip-gram model training phase. The entire model needs to be retrained if a new API call, which is not part of the training set, is encountered. 
 Scott et al. \cite{freitas2021large} proposed MalNet, a large scale Android malware FCG dataset, and they applied state-of-the-art graph representation learning approaches such as GIN \cite{xu2018powerful} for Android malware classification. Among all the methods, Feather and GIN achieved the highest classification performance.

In contrast, the proposed framework presented in this paper uses an inductive learning approach, which does not suffer from this limitation.

\section{\uppercase{Background}}
\subsection{Android FCG}
Overall, the APKs can be presented by the combination of its methods. Formally, an FCG is a directed graph $G=(N, E)$, where $N$ is a set of nodes representing Android API function, and $E$ represents the set of inter-procedural calls. FCGs can be very useful for Android malware classification. For example, when an app requires to send an SMS message, it has to perform a series of API calls on the Android platform. An FGC consists of all possible execution paths called during its runtime. For example, the app consists of the \textit{steal} function, which calls a list of functions to gather sensitive information such as phone contacts, SMS, the browser's bookmark, etc. Then, the sensitive information can be written into an XML file and sent back to the attacker by using \textit{sendData} functions.

\subsection{Graph Neural Networks}

Convolutional Neural Networks (CNNs) have been very successfully applied to the image classification problem. However, CNNs cannot be applied to non-Euclidean data structures. Therefore, GNNs can be thought of as a generalization of Convolutional Neural Networks to non-Euclidean data structures \cite{Bronstein}. GNNs have recently received a lot of attention due to their high interpretability via visualization of the graph \textit{embeddings} \cite{Zhou2018GraphNN}.

GNNs aim to generate node \textit{embeddings} \cite{Cai} which transform the graph nodes to a low-dimensional embedding space. All node embeddings can be passed through the readout function (i.e. via taking the mean of all node embeddings) to form the whole graph embeddings \cite{Cai}, which encode the whole graph into low-dimensional space for graph classification.

The FCG consists of structural information by modelling a set of functions and inter-procedural calls. The objects are represented by graph nodes and their relationships by graph edges. As a result, we can use GNNs for Android malware classification. In this paper, we evaluated three variants of GNNs for malware detection and classification, which are described below.

\subsubsection{\textbf{ Graph Convolutional Networks}}\label{GCN}
GCN is the most representative GNN to compute node embeddings by aggregating neighbour nodes' features. We consider graph $G = (N,E,A)$, where $N$ represents the set of nodes and $E$ the set of edges. $|N|$ is the number of nodes in the graph, and $|E|$ is the number of edges. The adjacency matrix $A$ is an $N \times N$ sparse matrix with $(i,j)$. Each node has a k-dimensional feature vector, and $X \in \mathbb{R}^{N \times K}$ represent the feature matrix for all $N$ nodes. An L-layer GCN \cite{kipf2016semi} consists of $L$ graph convolution layers, and each of them constructs embeddings for each node by mixing the embeddings of the node's neighbours in the graph from the previous layer.

\begin{equation}
Z^{l+1}=\sigma\left(X^{l} W_{0}^{l}+\tilde{A} X^{l} W_{1}^{l}\right)
\end{equation}

where $X^{(l)} \in \mathbb{R}^{N \times K_{l}}$ is the embedding at the $l$-th layer for all the N nodes and $X^{(0)}=X$. $W^{(l)}$ is the weight matrix that will be learnt for the downstream tasks. The $\sigma$ is an activation function that is usually set to be the element-wise ReLU. Let there be $L$ layers of graph convolutions, the output $Z(L)$ is the matrix consists of all node embeddings after $L$ layer transformations.

\subsubsection{\textbf {GraphSAGE}}\label{GraphSAGE Algorithm}
The \textit{Graph SAmple and aggreGatE (GraphSAGE)} algorithm was developed by Hamilton et al. \cite{Hamilton2017}. 
In GraphSAGE, unlike GCNs, a fixed size sub-set of node neighbours are (uniformly randomly) sampled. This allows limiting the space and time complexity of the algorithm, irrespective of the graph structure and batch size. Similar to the convolution operation in CNNs,  information relating to a node's local neighbourhood is collected and used to compute the node embedding.

At each iteration, the node's neighbourhood is initially sampled, and the information from the sampled nodes is aggregated into a single vector. 
At the $k$-th layer, the aggregated information $\mathbf{h}_{N(v)}^{k}$ at a node $v$, based of the sampled neighborhood $N(v)$, can be expressed as follows:  

\begin{equation}
\label{eq:general_aggregator}
    \mathbf{h}_{\mathcal{N}(v)}^{k} = \text { AGG }_{k}\left(\left\{\mathbf{h}_{u}^{k-1}, \forall {u} \in \mathcal{N}(v)\right\}\right)\
\end{equation}

Here, $\mathbf{h}_{u}^{k-1}$ represents the embedding of node $u$ in the previous layer. These embeddings of all nodes $u$ in the neighbourhood of $v$ are aggregated into the embedding of node $v$ at layer $k$. The aggregators  $AGG$ can be implemented as a \textit{mean, pooling or LSTM aggregator} function.

The aggregated embeddings of the sampled neighbourhood $\mathbf{h}_{N(v)}^{k}$ are then concatenated with the node's embedding from the previous layer $\mathbf{h}_v^{k-1}$. 
After applying the model's trainable parameters ($\mathbf{W}^k$, the trainable weight matrix) and passing the result through a non-linear activation function $\sigma$ (e.g. ReLU), the layer $k$ node $v$ embedding is calculated, as shown in Equation \ref{eq:general_graphsage}.
% e generalisation of the aggregation formula in GraphSAGE can be stated by Equation \ref{eq:general_graphsage} 

 \begin{equation}
 \small
 \label{eq:general_graphsage}
 \mathbf{h}_{v}^{k} = \sigma\left(\mathbf{W}^{k}\cdot\operatorname{CONCAT}\left(\mathbf{h}_{v}^{k-1}, \mathbf{h}_{\mathcal{N}(v)}^{k}\right)\right) 
\end{equation}

The final representation (embedding) of node $v$ is expressed as $\mathbf{z}_v$, which is essentially the embedding of the node at the final layer $K$, as shown in Equation \ref{eq:final_embedding}.
For the purpose of node classification, $\mathbf{z}_v$ can be passed through a sigmoid neuron or softmax layer.

 \begin{equation}
 \label{eq:final_embedding}
     \mathbf{z}_v = \mathbf{h}_v^K,\ \ \ \ \ \forall v \in \mathcal{V}
 \end{equation}
%  where $K$ is the k-hop neighbour or number of graph convolutional layers. 
%  \vspace{0.5cm}
 
%  \item \textbf{Back Propagation}

\subsubsection{\textbf{Graph Isomorphism Network}}\label{GIN}

Graph Isomorphism Network (GIN) was proposed by Xu et al. \cite{xu2018powerful}. The main difference between GIN and other GNNs is the message aggregation function part which is shown below:
\begin{equation}
h_{v}^{(k)}=\mathrm{MLP}^{(k)}\left(\left(1+\epsilon^{(k)}\right) \cdot h_{v}^{(k-1)}+\sum_{u \in \mathcal{N}(v)} h_{u}^{(k-1)}\right)
\label{eq:gin}
\end{equation}

Message passing of traditional GNNs is less powerful than the  Weisfeiler-Lehman (1-WL) \cite{shervashidze2011weisfeiler} algorithm. As the aggregation functions of the GNNs can be the same as the hash function of the 1-WL algorithm. Thus, update functions are not necessarily injective. Therefore, GIN \cite{xu2018powerful} was proposed to make the aggregation function to be injective, as shown in Equation \ref{eq:gin}, where $\varepsilon^{(k)}$ is a scalar parameter and MLP stands for a multi-layer perceptron.

 \begin{figure}[!t]
    \centering
        \includegraphics[width=1.0\columnwidth]{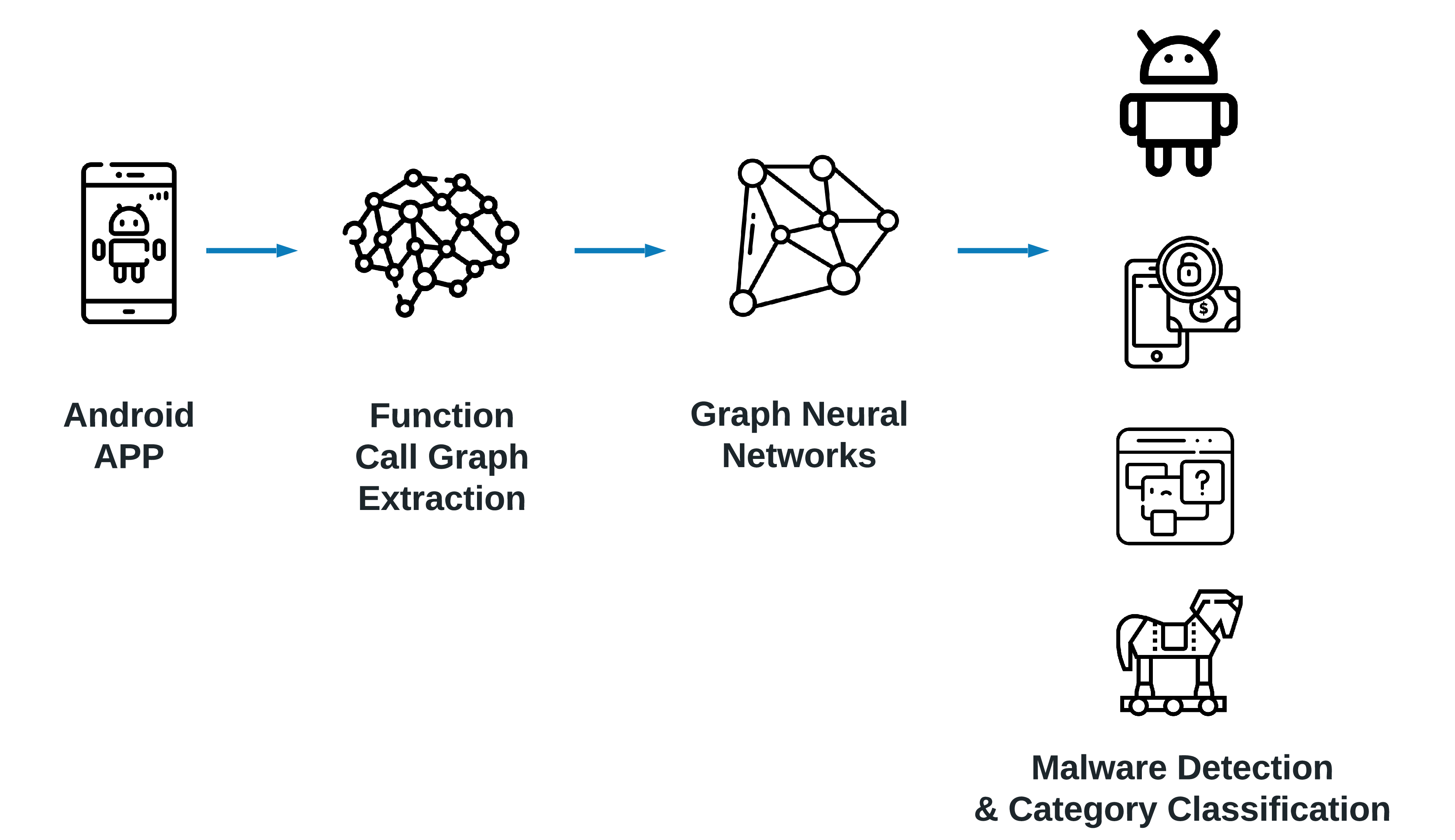}
        %\label{rfidtest_yaxis}
    \caption{Proposed Framework}
    \label{fig:Proposed Framework}
\end{figure}

\section{Proposed Framework}\label{Proposed}
Figure~\ref{fig:Proposed Framework} shows a high-level overview of our proposed framework for Android malware classification and analysis. 
%
% of our proposed E-GraphSAGE-based NIDS architecture. Although, there are data preprocessing stages such as data cleaning, formatting and standardisation, that is general in all ML-based NIDSs and have not been shown in the architecture. 
%
First, the Android malware function graphs are extracted from the Android APK's DEX file and fed into different GNN models for supervised FCG classification training. After the training process, the GNN models can perform malware detection, type and family classification based on FCGs. These three steps are explained in the following subsections. 

\subsubsection{FCG Extraction}\label{FCG Extraction}
FCGs consist of rich function graph structures to perform Android malware classification, which we aim to exploit in our approach. We convert the APK files to FCGs using Androguard \cite{desnos2013androguard}, which statically analyzes the APK's DEX file to generate FCGs. After the extraction process, the FCGs are used to train the GNN models for Android malware categories classification.

The FCGs are featureless, i.e. they do not contain of any node or edge features. In this paper, we perform graph node feature initialization based on the following graph centralities inspired by \cite{cui2021positional}. The graph centralities are standardized as the node features for function graph classification.

\begin{enumerate}
  \item \textbf{PageRank}: PageRank \cite{page1999pagerank} was proposed in \cite{page1999pagerank} and has been successfully applied to ranking web-pages based on graph-based techniques. The key idea behind PageRank is that highly linked pages are more influential than pages with few backlinks, and pages linked by highly influential pages are more important than pages linked by less important pages. Based on this idea, we adopt PageRank to calculate the importance factor of each function in the FCGs. The PageRank of a graph node $n_{i}$ is defined as fellow:
  
\begin{equation}
    P R\left(n_{i}\right)=\alpha \sum_{j \in B\left(n_{i}\right)} \frac{P R\left(n_{j}\right)}{\left|F\left(n_{j}\right)\right|}+\frac{1-\alpha}{N}
\end{equation}

 where $\alpha$ is the damping factor; $N$, the total number of function nodes; $B\left(n_{i}\right)$, the set of function nodes that links to $n_{i}$; and $\mid F\left(n_{j}\right)\mid$ is the number of forward links on node $n_{j} .$ In this paper, we apply dumping factor of 0.85 as it is a usually setting \cite{esuli2007pageranking}. 
  
  \item \textbf{In/out degree}: The in/out degree of each function nodes in FCGs.
  \item \textbf{Node betweenness centrality}: Node betweenness centrality \cite{brandes2001faster} measures the number of times a  function node lies on the shortest path between other function nodes. The betweenness centrality of a function node $v$ is defined as fellow:
  \begin{equation}
        c_{B}(v)=\sum_{s, t \in V} \frac{\sigma(s, t \mid v)}{\sigma(s, t)}
\end{equation}

where $V$ is the set of function nodes, $\sigma(s, t)$ is the number of shortest $(s, t)$-paths, and $\sigma(s, t \mid v)$ is the number of those paths passing through some function node $v$ other than $s, t$.\newline

\end{enumerate}

\subsubsection{Graph Neural Network Training}\label{Graph Neural Network Training}

\begin{figure}[!h]
    \centering
 \includegraphics[width=6.3cm,height=8cm]{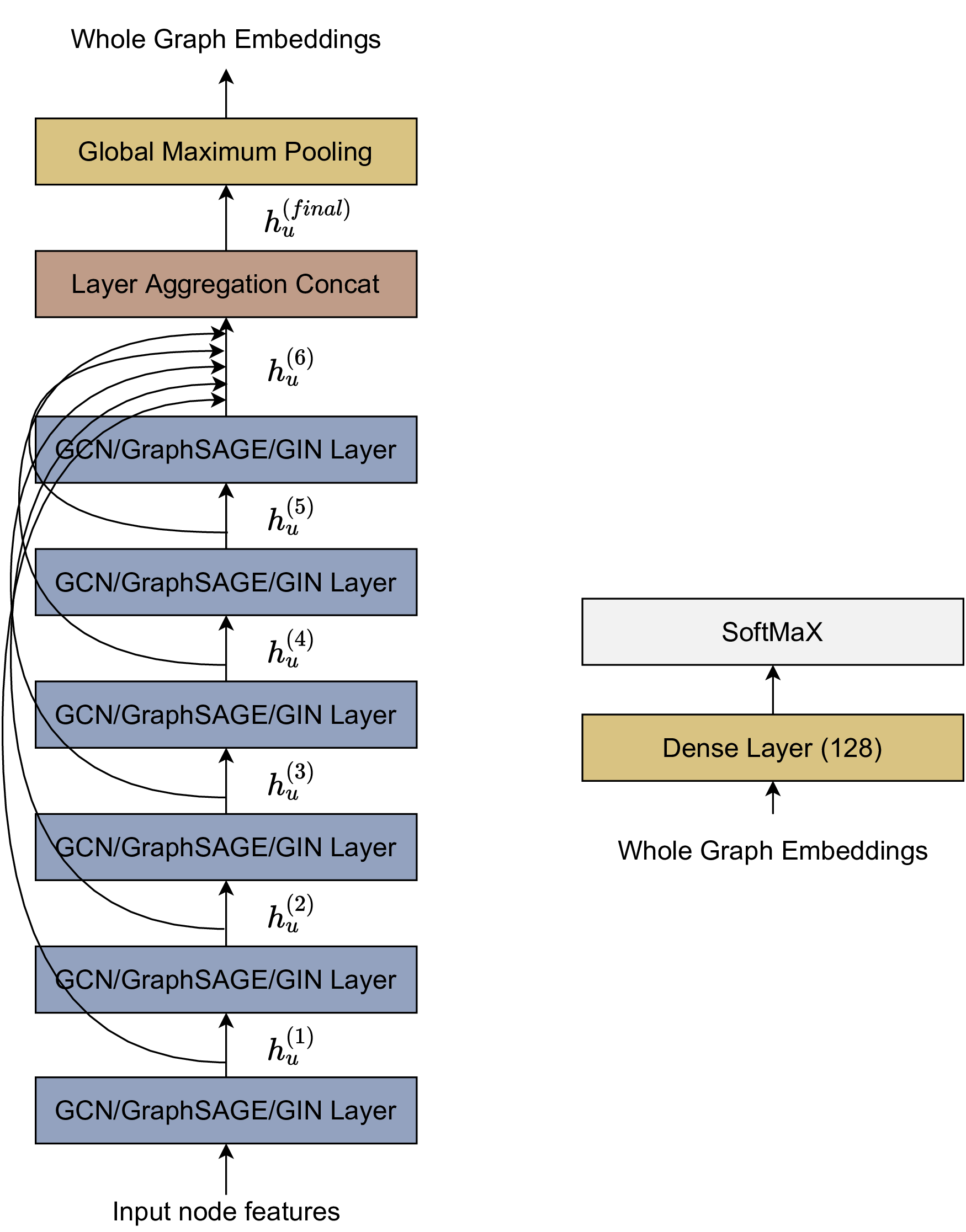}
        %\label{rfidtest_yaxis}
    \caption{Proposed Graph Neural JK Network Architecture}
    \label{fig:JK}
\end{figure}

In this paper, we evaluate three powerful and effective GNNs in conjunction with the JK technique \cite{xu2018representation} for Android malware detection and category classification based on FCGs, as shown in Fig. \ref{fig:JK}. Similar to the convolutional neural network, GNN models of increasing depth perform worse\cite{li2019deepgcns}. This is mainly due to over-smoothing  problem \cite{li2019deepgcns}.

To mitigate the effects of over-smoothing, we applied the JK \cite{xu2018representation} technique by using the concatenation layer. The key idea of JK is to select from all of those intermediate node representations and jump to the last layer for combining the intermediate node representations, to generate the final node representation. In this paper, we applied Layer aggregation concatenation to combine all intermediate node representations $\left(h_{u}^{(1)}, \ldots, h_{u}^{(k)}\right)$ for linear transformation to compute the final node embeddings $h_{u}^{(final)}$. 

The final node representation undergoes global-maximum pooling, which performs element-wise max-pooling over the final node embeddings $h_{u}^{(final)}$ for calculating the whole graph embeddings $r^{(i)}$ for function graph classification. In the following, we describe the model architecture and relevant hyperparameter settings.

\begin{enumerate}
 \item \textbf{GCN-JK}: GCN is the most representative graph representation approach, which obtains the node embedding by aggregating the neighbour nodes’ features. In this paper, we tried a 6 GCN layer JK network with $\in\{64, 128\}$ hidden units to compute the node embedding, and applied max-pooling over the node embeddings in a graph to compute the graph embedding. We found that a 6 layer GCN JK network with 128 hidden units performs the best.
 
  \item \textbf{GraphSAGE-JK}: GraphSAGE is another recent and relevant GNN method. Unlike GCN, which considers the entire set of neighbour nodes to obtain the node embedding, GraphSAGE uses a fixed set of neighbour nodes to reduce memory requirements. In this paper, we tried a 6 GraphSAGE max-pooling layer JK network with $\in\{64, 128\}$ hidden units to compute the node embedding, and we applied max-pooling over the node embeddings in a graph to compute a whole graph embedding. We found that a 6 layer GraphSAGE max-pooling JK network with 128 hidden units performs the best.
  
    \item \textbf{GIN-JK}: GIN is the most powerful GNN method. Traditional GNNs are less effective than the 1-WL algorithm \cite{shervashidze2011weisfeiler} due to their injective nature. Therefore, GIN was proposed to make the aggregation function injective.   In this paper, we used a 6 layer GIN JK network with $\in\{64, 128\}$ hidden units and $\epsilon=0$ to compute the node embedding,  and we applied max-pooling over the node embeddings in a graph to compute a whole graph embedding. We found that a 6 layer GIN JK network with 128 hidden units performs best.
\end{enumerate}

After the graph embedding computation, the graph embedding can be passed through a dense layer with 128 units and a ReLU activation function,  followed by a softmax layer for malware detection and classification. In this paper, we used the Adam optimizer with a learning rate of $\in\{0.001, 0.0001\}$, which provided the best results. 

\section{Experimental Results}\label{Results}
For evaluating the performance of the different GNN models, the standard metrics listed in Table~\ref{tab: metrics} are used, where $TP$, $TN$, $FP$ and $FN$ represent the number of True Positives, True Negatives, False Positives and False Negatives, respectively.

\subsection{Datasets}\label{Datasets}

For the evaluation, we used two relevant publicly available Android malware datasets, consisting of different types of malware categories/families. 

%TTTTTTTTTTTTTTTTTTTTTTTTTTTTTTTTTTTTTTTTTTT
\begin{table}[!t]
\renewcommand{\arraystretch}{1.4}
\caption{Evaluation metrics utilised in this study.}
\centering
\begin{tabular}
{cc} \hline
\textbf{Metric} & \textbf{Definition} \\ \hline 
\small{ Recall (Detection Rate)} &  $\frac{TP}{TP+FN}$ \\  \hline
\small{Precision} & $\frac{TP}{TP+FP}$ \\ \hline
\small{F1-Score} & $ 2 \times \frac{Recall\times Precision}{Recall + Precision}$ \\ \hline
\small{Accuracy} &   $\frac{TP+TN}{TP+FP+TN+FN}$ \\ \hline

\end{tabular}
\label{tab: metrics}
\end{table}
%TTTTTTTTTTTTTTTTTTTTTTTTTTTTTTTTTTTTTTT

\begin{enumerate}

\item \textbf{Malnet-Tiny}: Malnet-Tiny is an Android malware FCG dataset that was created by Scott et al. \cite{freitas2021large} from Georgia Tech University and the Microsoft APT team. The dataset consists of 4,500 malicious FCGs, belonging to four different malware categories, and 500 benign FCGs. We follow the approach for training, validation, and testing split (70/10/20), as specified by the authors. 

\item \textbf{Drebin}: Drebin \cite{arp2014drebin} is an Android malware dataset that includes 5,560 APKs from 179 different malware families. The dataset was collected during the period of August 2010 to October 2012. Since Drebin only consists of malware samples, we also download 12,686 benign APK files from AndroZoo \cite{allix2016androzoo}, a large-scale APK repository, with samples collected from 2013 to 2019 for the malware detection experiments. The dataset only consists of malware family labels rather than category labels. Thus, we only perform malware family classification. We list the top 24 malware families and perform malware family classification for those top families. For these experiments, we randomly split the combined and Drebin datasets into 70\% for training and 30\% for testing. 
\end{enumerate}

\subsection{Malnet-Tiny Classification Results}\label{Binary Classification Results}

% TTTTTTTTTTTTTTTTTTTTTTTTTTTTTTTTTTTTTTTT
\begin{center}
\begin{table}[H]\small
\caption{Malnet-Tiny Multiclass classification results.}
\resizebox{\columnwidth}{!}{\begin{tabular}{cccccccc}
\hline \textbf{Method} & \textbf{Accuracy} & \textbf{Precision} & \textbf{F1-Score} & \textbf{Recall}  %& Prediction Time (µs) 
\\   \hline

GCN-JK & 89.70\% & 89.87\% & 0.90 & 89.70\% 
\\
 GraphSAGE-JK & 94.40\% & 94.53\% & 0.94 & 94.40\%  
\\
 GIN-JK & 90.00\% & 90.59\% & 0.90 & 90.00\%  
\\
\hline
\end{tabular}
}
\label{tab:malnet_jk}
\end{table}
\end{center}
%TTTTTTTTTTTTTTTTTTTTTTTTTTTTTTTTTTTTTTTTT

%TTTTTTTTTTTTTTTTTTTTTTTTTTTTTTTTTTTTTTTTTT
\begin{table}[h!]\small
\centering
\caption{Results of multiclass classification by GraphSAGE-JK on Malnet-Tiny dataset}
\label{tab:multi-bot-iot}
\begin{tabular}{rrrr}
\cline{2-3}
                                               & \multicolumn{2}{c}{\textbf{GraphSAGE-JK}}  \\ \hline
\multicolumn{1}{c}{\textbf{Class Name}} &
\multicolumn{1}{c}{\textbf{Recall}} &
\multicolumn{1}{c}{\textbf{F1-Score}} \\   \hline

\multicolumn{1}{l}{{Benign}}  & 91.00\%    & 0.88                    \\ 
\multicolumn{1}{l}{{AdDisplay}}        & 97.00\%      & 0.97    \\ 
\multicolumn{1}{l}{{Adware}}        & 95.00\%              & 0.95  \\ 
\multicolumn{1}{l}{{Downloader}}        & 99.50\%    & 0.99     \\ 
\multicolumn{1}{l}{{Trojan}}   & 89.50\% & 0.93 \\ \hline

\multicolumn{1}{l}{\textbf{Weighted Average}}   & \textbf{94.40\%}              & \textbf{0.94}  \\ \hline
\end{tabular}
\label{tab:malnet_graphsage}
\end{table}
%TTTTTTTTTTTTTTTTTTTTTTTTTTTTTTTTTTTTTTTTTT

As the Malnet-Tiny dataset only provides the multiclass splitting configuration, we only perform multiclass experiments in this case. Tables \ref{tab:malnet_jk} and \ref{tab:malnet_graphsage} show the corresponding multiclass results for Malnet-Tiny. In this experiment, GraphSAGE-JK performed the best. It achieves a very high Weighted-Recall and F1-Score across 5 application classes (4 malware types plus benign class), with a Weighted-Recall and Weighted-F1 score of 0.94. The GraphSAGE-JK model achieved the worst detection performance for the trojan malware class, with 89.50\% detection rate, and the best for the downloader class, with a detection rate of 99.50\%. 

%TTTTTTTTTTTTTTTTTTTTTTTTTTTTTTTTTTTTTTTTTTTTTTTTTTTTTTT
\begin{table}[h!]\small
\centering
\caption{Performance of Malnet-Tiny multiclass classification by JK-Networks compared with the baseline algorithms.}
\begin{tabular}{ccc}
%  \hline
%  \multicolumn{4}{|c|}{Multiclass Classification Performance Comparison with baselines Algorithms} \\
 \hline
  \textbf{Method}  &   \textbf{Accuracy}\\
  \hline
 Feather \cite{freitas2021large}     & 86.00\% \\
 LDP \cite{freitas2021large}    & 86.00\% \\
 GIN  \cite{freitas2021large}    & 90.00\% \\
 GCN \cite{freitas2021large}    & 81.00\% \\
 Slaq-LSD  \cite{freitas2021large}   & 76.00\% \\
 NoG  \cite{freitas2021large}  & 77.00\% \\
 Slaq-VNGE \cite{freitas2021large}   & 53.00\% \\ \hline
  GCN-JK   &   89.70\% \\
  GIN-JK    &   90.00\% \\
\textbf{  GraphSAGE-JK }  &   \textbf{94.40\%} \\

 \hline
\end{tabular}
\label{table:multi_comp}
\end{table}
%TTTTTTTTTTTTTTTTTTTTTTTTTTTTTTTTTTTTTTTTTTTTTTTTTTT

Table~\ref{table:multi_comp} shows the average multiclass accuracy of the considered JK-Networks compared to the baseline results from the literature, provided in \cite{freitas2021large}. Since the distribution across various application classes is balanced, we can use the accuracy as a meaningful performance metric. We observe that GraphSAGE-JK outperforms all baseline classifiers. Moreover, we observe that, compared with the original deeper GCN performance \cite{freitas2021large}, our JK-based GCN approach can improve the original GCN performance from 81.00\% to 89.70\%. 

\subsection{Drebin Classification Results}\label{Binary Classification Results}
We are now considering the Drebin dataset. For the evaluation of the Drebin experiments, we conducted both binary (malware/benign) and multiclass Android malware classification.   

\subsubsection{Binary Classification Results}
% TTTTTTTTTTTTTTTTTTTTTTTTTTTTTTTTTTTTTTTT
\begin{center}
\begin{table}[h!]
\caption{Drebin Binary classification results.}
\resizebox{\columnwidth}{!}{\begin{tabular}{cccccccc}
\hline \textbf{Method} & \textbf{Accuracy} & \textbf{Macro Precision} & \textbf{Macro F1} & \textbf{Macro Recall}  %& Prediction Time (µs) 
\\   \hline

 GCN-JK & 97.82\% & 97.06\% & 0.97 & 97.81\% 
\\
 GraphSAGE-JK & 98.00\% & 97.49\% & 0.98 & 97.77\%  
\\
 GIN-JK & 97.39\% & 96.86\% & 0.97 & 96.96\%  
\\
\hline
\end{tabular}
}
\label{table:drebin_binary}
\end{table}
\end{center}
%TTTTTTTTTTTTTTTTTTTTTTTTTTTTTTTTTTTTTTTTT
Tables \ref{table:drebin_binary} shows the corresponding malware detection results for the Drebin dataset. Across all methods, GraphSAGE-JK again performs the best. It achieves a high detection performance with a recall and F1 score of 97.77\% and 0.98, respectively.

\subsubsection{Multiclass Classification Results}

% TTTTTTTTTTTTTTTTTTTTTTTTTTTTTTTTTTTTTTTT
\begin{center}
\begin{table}[H]\small
\caption{Drebin Multiclass classification results.}
\resizebox{\columnwidth}{!}{\begin{tabular}{cccccccc}
\hline \textbf{Method} & \textbf{Accuracy} & \textbf{Precision} & \textbf{F1-Score} & \textbf{Recall}  %& Prediction Time (µs) 
\\   \hline

GCN-JK & 95.24\% & 95.35\% & 0.95 & 95.24\% 
\\
 GraphSAGE-JK & 96.88\% & 97.01\% & 0.97 & 96.88\%  
\\
 GIN-JK & 95.03\% & 95.07\% & 0.95 & 95.03\%  
\\
\hline
\end{tabular}
}
\label{tab:drebin-mul}
\end{table}
\end{center}
%TTTTTTTTTTTTTTTTTTTTTTTTTTTTTTTTTTTTTTTTT

\begin{figure*}[t]
     \centering
     \begin{subfigure}[b]{0.3\textwidth}
         \centering
         \includegraphics[width=1.0\textwidth]{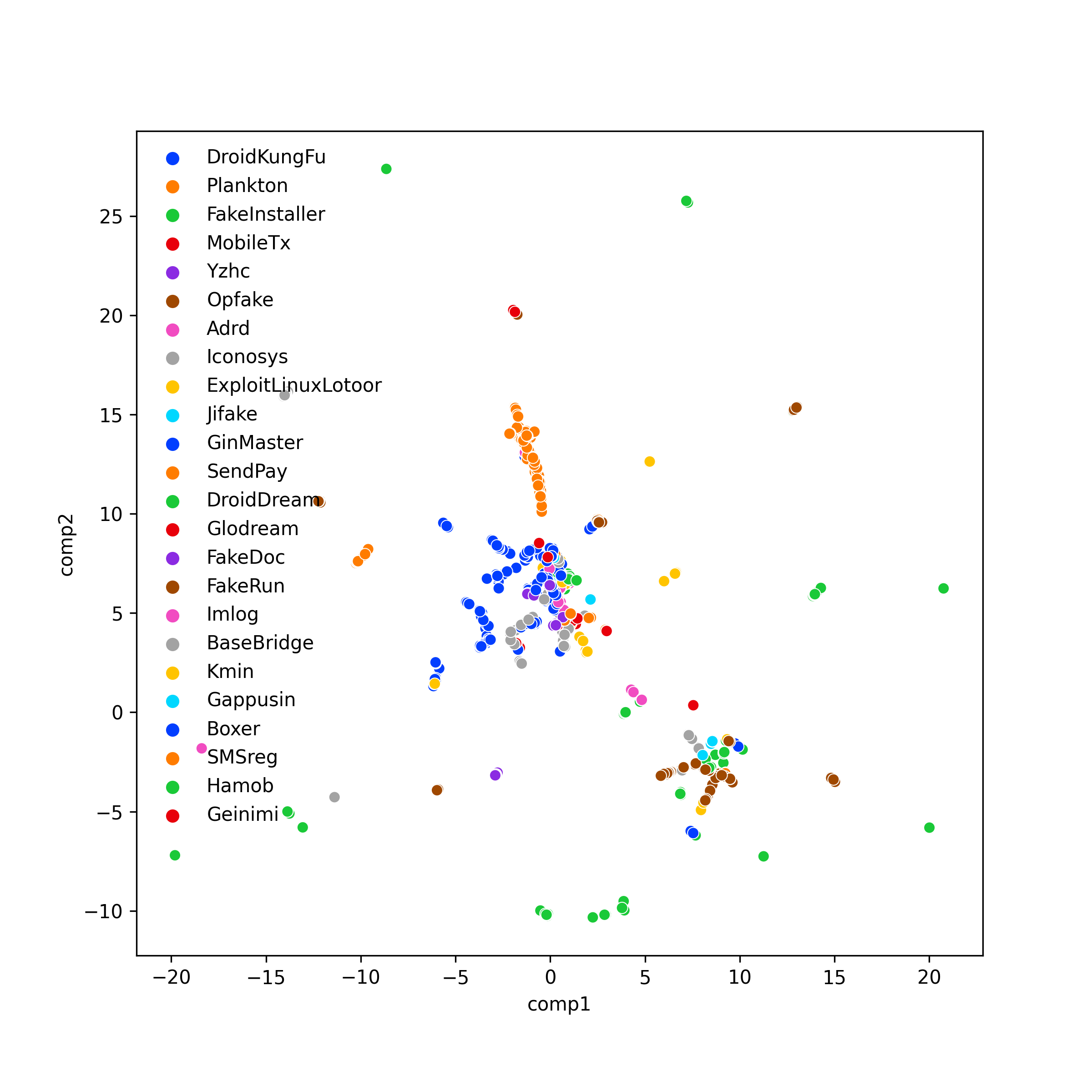}
         \caption{}
         \label{fig:y gcn_jk_drebin}
     \end{subfigure}
     \hfill
     \begin{subfigure}[b]{0.3\textwidth}
         \centering
         \includegraphics[width=1.0\textwidth]{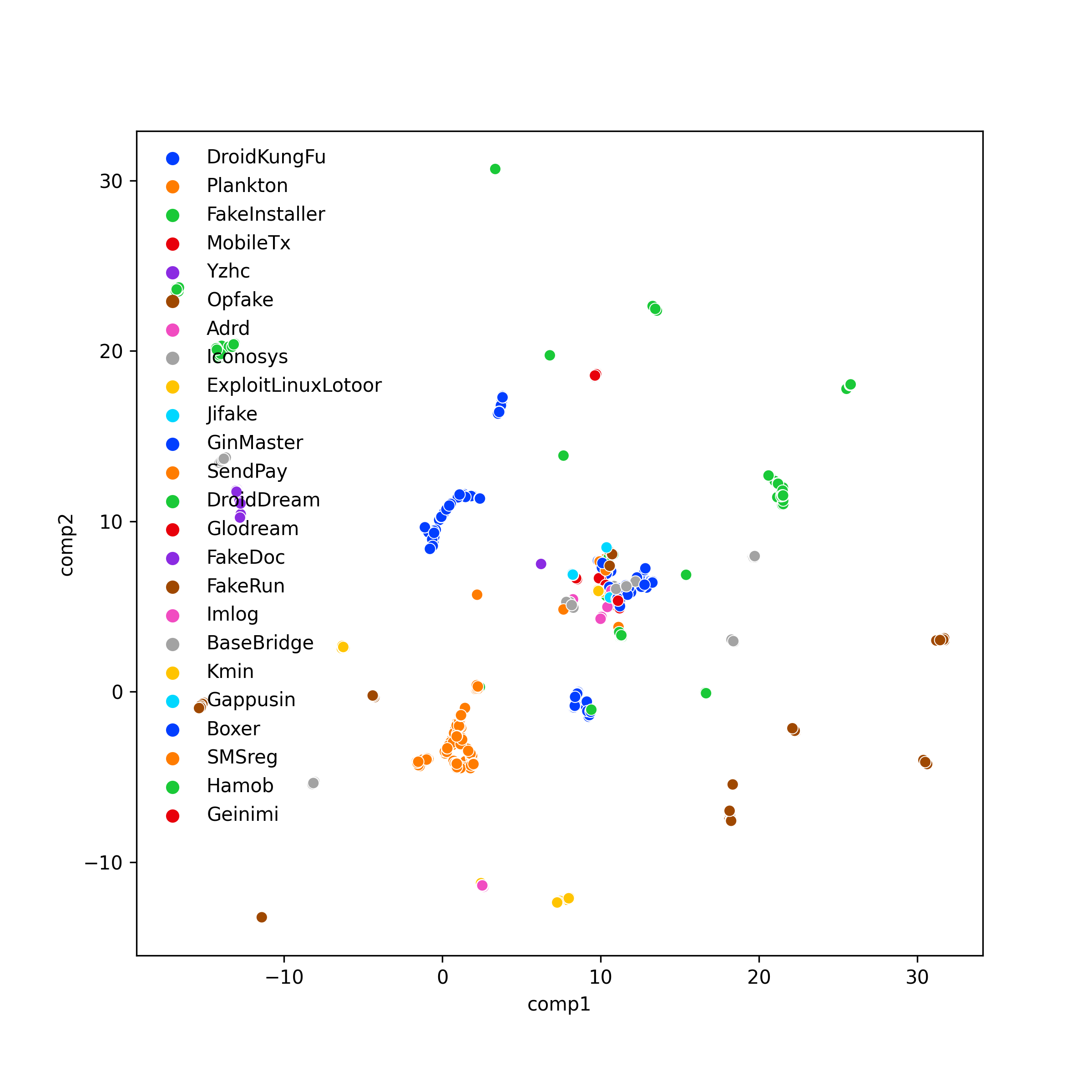}
         \caption{}
         \label{fig:graphsage_jk_drebin}
     \end{subfigure}
     \hfill
     \begin{subfigure}[b]{0.3\textwidth}
         \centering
         \includegraphics[width=1.0\textwidth]{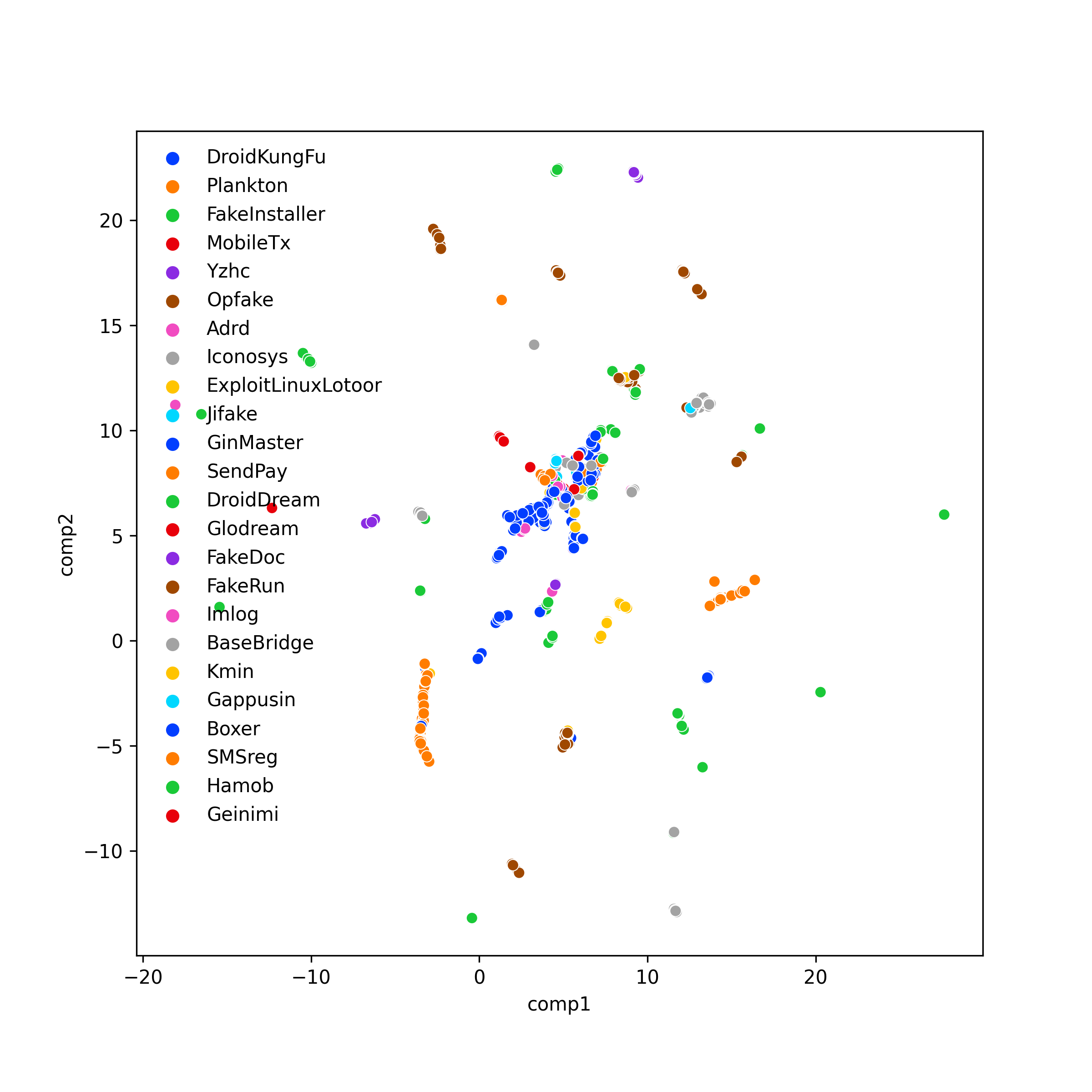}
         \caption{}
         \label{fig: gin_jk_drebin}
     \end{subfigure}
        \caption{UMAP visualizations of whole graph embeddings of all three models on the Drebin dataset. (a) GCN-JK (b) GraphSAGE-JK (c) GIN-JK }
        \label{fig:umap_drebin}
\end{figure*}

\begin{figure*}[t]
     \centering
     \begin{subfigure}[b]{0.3\textwidth}
         \centering
         \includegraphics[width=0.85\textwidth]{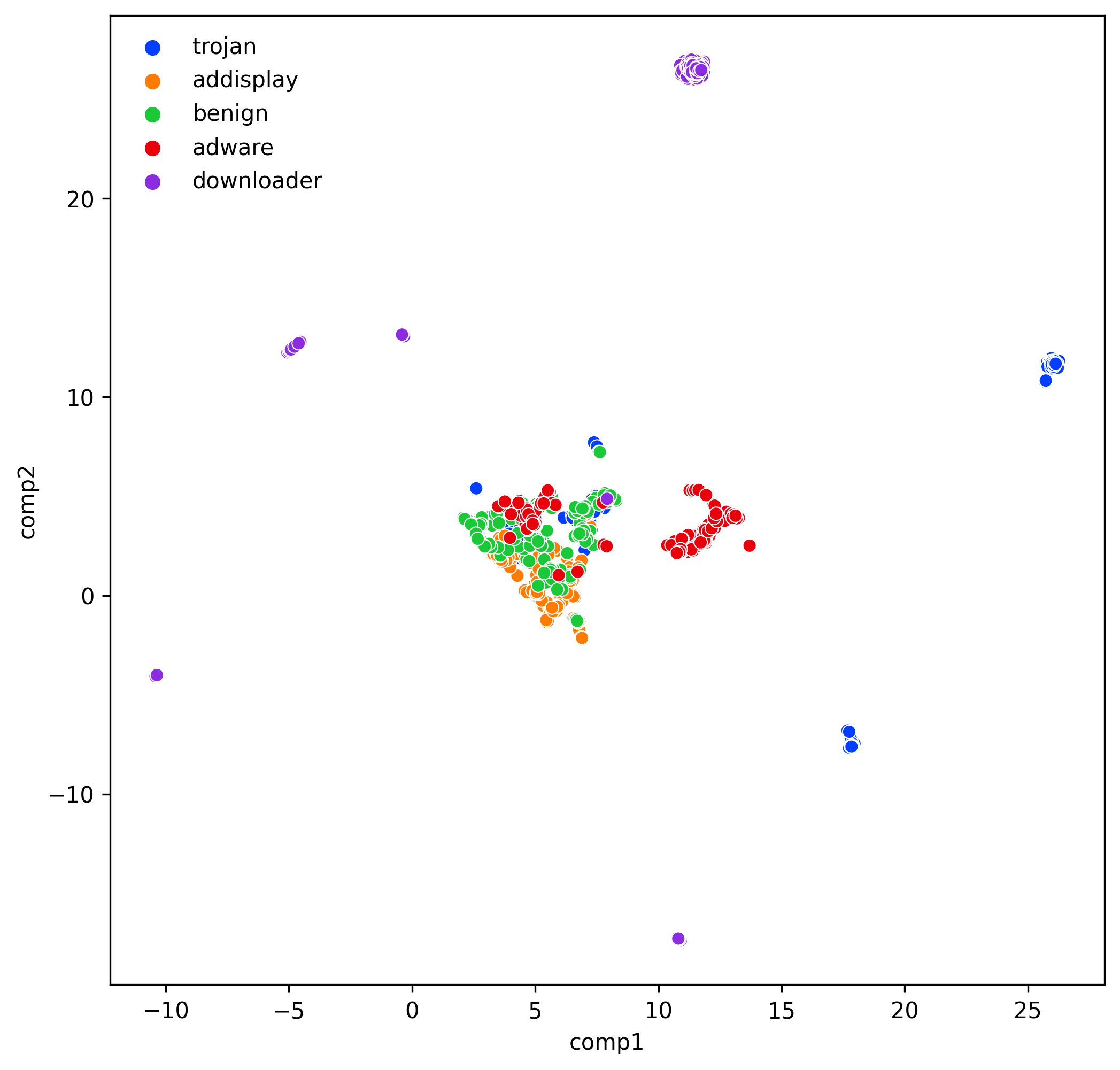}
         \caption{}
         \label{fig:y gcn_jk_malnet}
     \end{subfigure}
     \hfill
     \begin{subfigure}[b]{0.3\textwidth}
         \centering
         \includegraphics[width=0.85\textwidth]{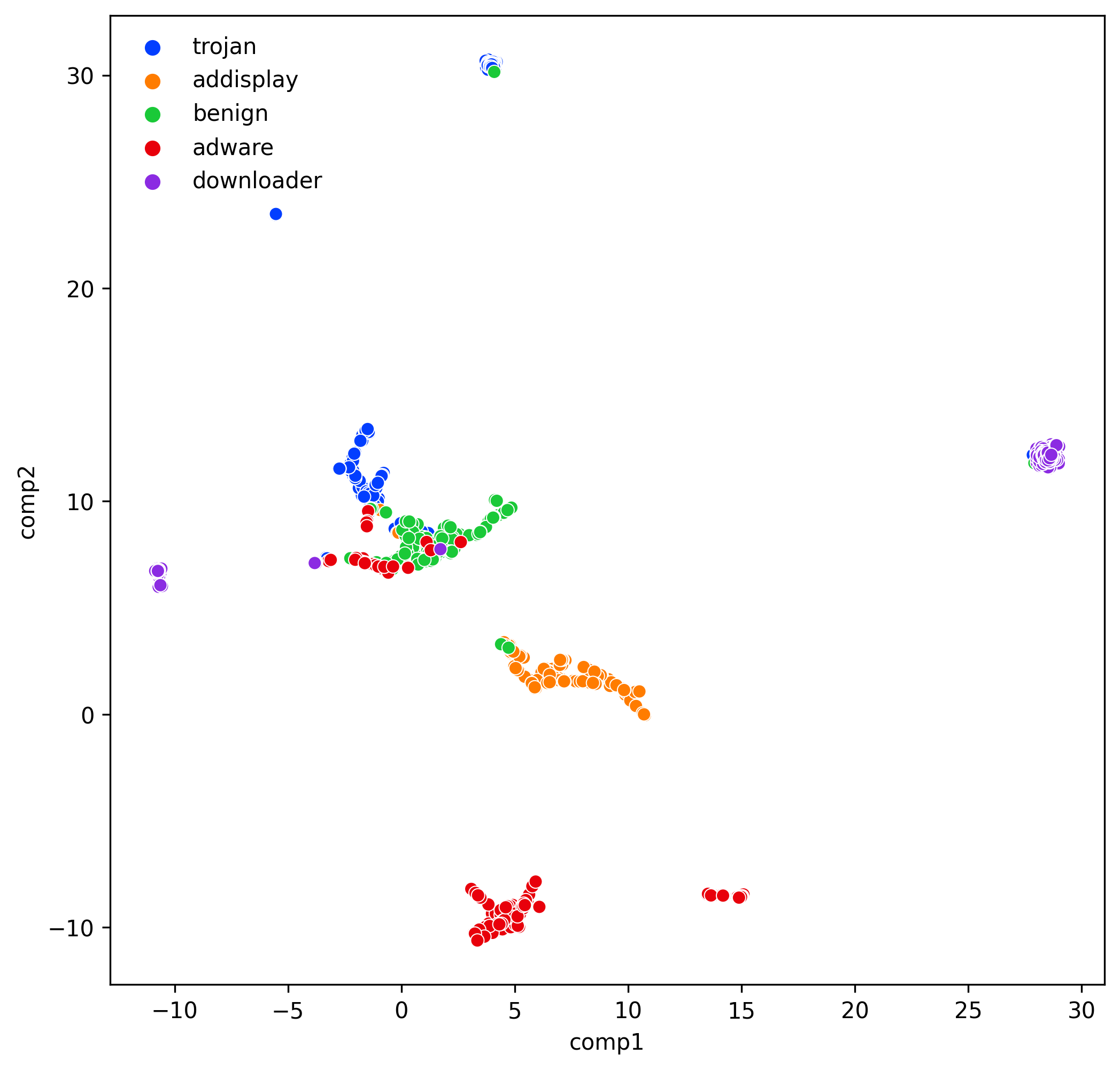}
         \caption{}
         \label{fig:graphsage_jk_malnet}
     \end{subfigure}
     \hfill
     \begin{subfigure}[b]{0.3\textwidth}
         \centering
         \includegraphics[width=0.85\textwidth]{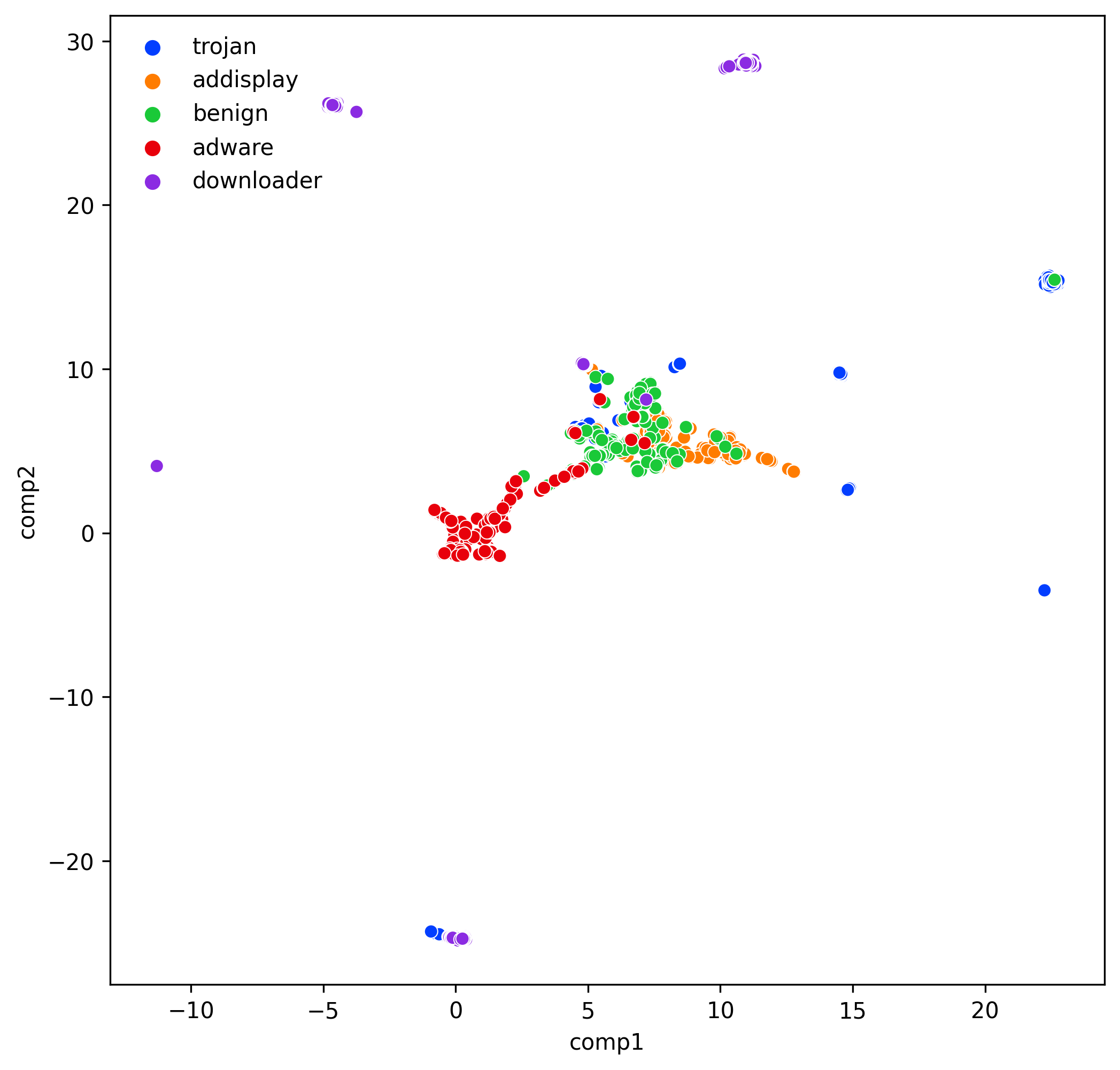}
         \caption{}
         \label{fig: gin_jk_malnet}
     \end{subfigure}
        \caption{UMAP visualizations of whole graph embeddings of all three models on the Malnet-tiny dataset. (a) GCN-JK (b) GraphSAGE-JK (c) GIN-JK }
        \label{fig:umap_malnet}
\end{figure*}

%TTTTTTTTTTTTTTTTTTTTTTTTTTTTTTTTTTTTTTTTTT
\begin{table}[h!]
\centering
\caption{Results of multiclass classification by GraphSAGE-JK on Drebin dataset}
\begin{tabular}{lrrrr}
\cline{2-3}
                                               & \multicolumn{2}{c}{\textbf{GraphSAGE-JK}}  \\ \hline
\multicolumn{1}{c}{\textbf{Class Name}} &
\multicolumn{1}{c}{\textbf{Recall}} &
\multicolumn{1}{c}{\textbf{F1-Score}} \\ \hline
\multicolumn{1}{l}{{Adrd}}  & 88.89\%    & 0.87                     \\
\multicolumn{1}{l}{{BaseBridge}}        & 88.89\%      & 0.93    \\
\multicolumn{1}{l}{{Boxer}}        & 100.0\%              & 1.00  \\
\multicolumn{1}{l}{{DroidDream}}        & 95.83\%    & 0.96     \\
\multicolumn{1}{l}{{DroidKungFu}}   & 98.99\% & 1.0 \\
\multicolumn{1}{l}{{ExploitLinuxLotoor}}   & 80.95\% & 0.72 \\
\multicolumn{1}{l}{{FakeDoc}}   & 100.0\% & 0.99 \\

\multicolumn{1}{l}{{FakeInstaller}}   & 98.20\% & 0.98 \\

\multicolumn{1}{l}{{FakeRun}}   & 100.0\% & 1.0 \\

\multicolumn{1}{l}{{Gappusin}}   & 100.00\% & 0.97 \\

\multicolumn{1}{l}{{Geinimi}}   & 100.0\% & 1.0 \\

\multicolumn{1}{l}{{GinMaster}}   & 88.24\% & 0.90 \\

\multicolumn{1}{l}{{Glodream}}   & 84.21\% & 0.84 \\

\multicolumn{1}{l}{{Hamob}}   & 100.0\% & 1.0 \\

\multicolumn{1}{l}{{Iconosys}}   & 100.0\% & 1.0 \\

\multicolumn{1}{l}{{Imlog}}   & 100.0\% & 0.96 \\

\multicolumn{1}{l}{{Jifake}}   & 88.89\% & 0.89 \\

\multicolumn{1}{l}{{Kmin}}   & 100.0\% & 1.0 \\

\multicolumn{1}{l}{{MobileTx}}   & 100.0\% & 1.0 \\

\multicolumn{1}{l}{{Opfake}}   & 99.46\% & 0.98 \\

\multicolumn{1}{l}{{Plankton}}   & 99.46\% & 0.99 \\

\multicolumn{1}{l}{{SMSreg}}   & 100.00\% & 0.92 \\

\multicolumn{1}{l}{{SendPay}}   & 100.0\% & 0.97 \\

\multicolumn{1}{l}{{Yzhc}}   & 100.0\% & 0.96 \\ \hline

\multicolumn{1}{l}{\textbf{Weighted Average}}   & \textbf{96.88\%}              & \textbf{0.97}  \\
\end{tabular}
\label{tab:drebin_graphsage}
\end{table}
%TTTTTTTTTTTTTTTTTTTTTTTTTTTTTTTTTTTTTTTTTT

Tables \ref{tab:drebin-mul} and \ref{tab:drebin_graphsage} show the corresponding multiclass results. Again, GraphSAGE-JK performed the best. It achieved a Weighted-Recall and F1-Score across 24 malware families of 96.68\% and 0.97, respectively. The classification performance of the top 24 malware families is shown in Table \ref{tab:drebin_graphsage}—all samples from the 24 largest malware families in the Drebin dataset.

Table~\ref{table:multi_comp_drebin} shows the average multiclass accuracy and detection rate of the three considered GNN JK-Networks compared with the state-of-the-art. We observe that GraphSAGE-JK outperforms most other approaches in terms of accuracy.
%TTTTTTTTTTTTTTTTTTTTTTTTTTTTTTTTTTTTTTTTTTTTTTTTTTTTTTT
\begin{table}[H]\small
\centering
\caption{Performance of Drebin malware detection by JK-Networks compared with the baseline algorithms.}
\begin{tabular}{ccc}
%  \hline
%  \multicolumn{4}{|c|}{Multiclass Classification Performance Comparison with baselines Algorithms} \\
   \hline

  \textbf{Method}  &   \textbf{Accuracy} &   \textbf{Detection Rate}\\
  \hline
 Drebin \cite{arp2014drebin}     & 93.90\% &  94.00\% \\

ICC \cite{xu2016iccdetector} & 97.40\% & 93.10\% \\ 

 FAMD \cite{bai2020famd}   &   97.40\%  &   96.77\% \\ 

\hline

 GCN-JK  &   97.82\% &  97.81\%  \\
  GIN-JK    &   97.39\% & 95.86\%  \\
\textbf{  GraphSAGE-JK }  &   \textbf{98.00\%}  & \textbf{97.20\%} \\

 \hline
\end{tabular}
\label{table:multi_comp_drebin}
\end{table}
%TTTTTTTTTTTTTTTTTTTTTTTTTTTTTTTTTTTTTTTTTTTTTTTTTTT

\section{Learned Feature Representation}\label{Representation}
Finally, we provide a visualization of the graph embeddings of all three models after a dense layer non-linear transformation by using the UMAP \cite{umap} dimensionality reduction. Figure~\ref{fig:umap_drebin} shows the UMAP results of the Drebin dataset. There is no significant difference in class separation among all three models. The visualization results of the Malnet-Tiny dataset are shown in Figure~\ref{fig:umap_malnet}. It is clear that GCN-JK and GIN-JK (in Figure~\ref{fig:y gcn_jk_malnet} and \ref{fig: gin_jk_malnet}) learn similar representations on Malnet-tiny. While these models perform well in the detection classes adware and downloader, they are less effective in distinguishing the other classes. Moreover, in terms of GraphSAGE-JK (in Figure \ref{fig:graphsage_jk_malnet}), comparing between two other pairs of models, we can see that a higher class separability can be achieved.  

\section{Limitations and Future work}\label{Limitations}
The model architecture deployed in the GNN-JK model is limited to reasonably complex 6-layer JK networks. Since the motivation of this paper is to investigate the effectiveness of FCGs and GNNs for Android malware detection and classification, the model we used was not optimized at all. In future works, it would be interesting to explore other GNN architectures, such as DeeperGCN \cite{li2020deepergcn}, EdgePool \cite{diehl2019edge}.

Moreover, some existing detection approaches, e.g. \cite{taheri2019extensible}, also try to use network flow for Android malware detection. However, they only consider flow data records independently and fail to detect sophisticated malware (i.e., the botnet malwares try to launch DDoS attacks to the victims) as a more global view of the network and traffic flow is required. As a result, it is worth exploring the edge-based graph neural approaches such as E-GraphSAGE \cite{lo2021graphsage}, E-ResGAT \cite{chang2021graph} to perform Android malware detection based on malicious network flows and combine with GNN-Based FCG's approach for Android malware detection.

Furthermore, it is worth exploring explainable graph neural network algorithms, such as SubgraphX \cite{subgraphx}, as it can help researchers analyze and explain the working process of GNNs to detect malware by highlighting suspicious function call paths for automatic malware forensics.

\section{Conclusion}\label{Conclusions}
This paper presents a novel approach for automatic malware detection and analysis based on GNNs with JK. We first used the JK network to detect Android malware with high detection rates effectively. Our experimental evaluation based on two benchmark datasets shows that our approach performs exceptionally well and overall outperforms the baseline ML-based/Graph-based Android malware classifier. The evaluation results of our initial classifier demonstrate the potential of a GNN-based approach for Android malware detection and classification. 

%%%%%%%%%%%%%%%%%%%%%

\printbibliography

\end{document}